\documentclass{ws-procs10x7}
\usepackage{balance}

\newcolumntype{d}[1]{D{.}{.}{#1}}

\makeindex
\begin{document}

\title{EXOTIC PHYSICS AT HERA}

\author{G. BRANDT}

\address{
\emph{(on behalf of the H1 and ZEUS collaborations)}\\
Physical Institute, University of Heidelberg, 
Philosophenweg 12, 69120 Heidelberg, Germany\\
E-mail: gbrandt@mail.desy.de}



\twocolumn[\maketitle\abstract{
A search for excited neutrinos, the analysis of multi-lepton final states, 
a search for doubly charged Higgs production and a general search for 
high-$P_T$ phenomena at HERA are reported.
The searches use data samples of $e^{\pm}p$ collisions with a centre-of-mass
energy $\sqrt{s}=320$ GeV collected
by the H1 and ZEUS experiments at HERA in the years 1994-2005 with integrated
luminosities up to $\mathcal{L}=296$ pb$^{-1}$. 
Overall no significant deviations of the experimental observations from the
Standard Model (SM) expectation are found.
}
\keywords{Search; Leptons; Neutrino; Compositeness; Higgs; BSM.}
]

\section{Introduction}

HERA is a collider on the energy frontier providing $ep$ collisions at a 
centre-of-mass energy $\sqrt{s}=320$ GeV. It can explore parts of the phase
space not reachable by other colliders, opening a unique window for the search
for new physics.
A classic search for physics beyond the Standard Model (BSM) is the search
for compositeness of elementary particles. Section \ref{sec:nustar} 
presents a search for excited neutrinos. 
It profits from the tenfold increase in statistics of the $e^{-}p$
data sample collected in HERA-II (2003-05, $\mathcal{L}\simeq180$ pb$^{-1}$) with 
respect to HERA-I (94-00, $\mathcal{L}\simeq20$ pb$^{-1}$).
This $e^{-}p$ data sample is also searched for events with multiple leptons in
the final state (Section \ref{sec:mlep}). 
Pairproduction of $\tau$-leptons is measured for the first time at HERA.
Of particular interest was the high mass and $P_T$ region in light of the 
excess of $ee$ and $eee$ events observed in the HERA-I sample. 
The significance of the excess does not increase using the full sample.
The interpretation of these events in terms of doubly charged Higgs $H^{\pm\pm}$
production is performed in a dedicated search (Section \ref{sec:hpp}). 
$H^{\pm\pm}$ production is ruled out for the $ee$-case.
Also a general, model-independent search for high-$P_T$ phenomena is performed on 
the HERA-II $e^{-}p$ data sample looking for deviations from the SM in a multitude
of final states with high-$P_T$ objects (Section \ref{sec:generic}).

\section{Search for Excited Neutrinos} \label{sec:nustar}

Compositeness of fermions provides a natural solution for the fermion mass hierarchy
problem in the SM\cite{hg}.
The excitation part of a Lagrangian for a minimal extension of the SM to allow excited fermions considering only electroweak interactions can be written
\begin{equation*}
L_{F^{*}F} = \frac{1}{2\Lambda}{\overline{F^{*}_{R}}}{{\sigma}^{\mu\nu}}[gf\frac{\overrightarrow{\tau}}{2}{\partial}_{\mu}{\overrightarrow{W_{\nu}}}+
\end{equation*}
\begin{equation*}
\qquad g'f'\frac{Y}{2}{\partial}_{\mu}B_{\nu}]{F_{L}} + {\mathrm h. c.} ,
\label{lg}
\end{equation*}
where the weights $f$ and $f'$ modifiy the electroweak SU(2) $\times$ U(1) couplings $g$ and $g'$, respectively. $W$ and $B$ denote the gauge boson fields,
$\sigma_{\mu\nu}$ the Pauli matrices and $Y$ the weak hypercharge. Compositeness
comes into the play at the scale $\Lambda$ and the weights $f$ and $f'$ determine
the production cross section and decay branching ratios of excited fermions.
The cross section for excited neutrino production is about 80 times larger in $e^{-}p$ than in $e^{+}p$ for high $M_{\nu^{*}}$ 
due to the $W$ exchange involved. 
 
\begin{figure}[thb]
\setlength{\unitlength}{1cm}
\psfig{file=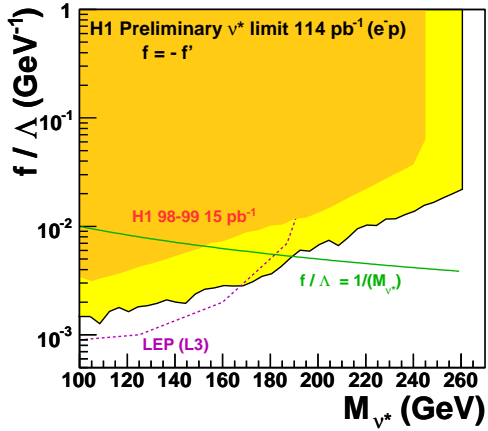,width=70mm}
\caption{Limits on the ratio $f/\Lambda$ for the assumption $f=-f'$ (maximal coupling $\gamma\nu\nu*$).}
\label{fig:nu_star}
\end{figure}

Investigated final states are
$\nu^{*}\rightarrow\nu\gamma$, 
$\nu^{*}\rightarrow\nu Z \hookrightarrow q \bar{q}$ and 
$\nu^{*}\rightarrow e W \hookrightarrow q \bar{q}$.
No signal of $\nu^{*}$ production was found\cite{nustar06-prel}. 
Limits on the cross section
in the plane of the ratio $f/\Lambda$ and the mass of the $\nu^{*}$, $M_{\nu^{*}}$
are presented in Fig. \ref{fig:nu_star} for the hypothesis $f=-f'$ corresponding
to maximal $\gamma\nu\nu^{*}$ coupling. 
Assuming $f/\Lambda = 1/M_{\nu*}$, excited neutrinos with masses below 188 GeV are excluded at 95\% CL.
The results extend previous searches at HERA \cite{Adloff:2001me} and are 
sensitive beyond LEP reach.

\section{Multi-Lepton Events} \label{sec:mlep}

Within the SM framework multi-lepton events at high $P_T$ are produced mainly via
$\gamma$-$\gamma$ interactions\cite{ref:ggtheo}.

An H1 preliminary analysis\cite{ref:h1-mlep} using a data sample corresponding to
275 pb$^{-1}$ selects two central leptons (electrons and muons) 
with $P_{T 1,2} > 10, 5$ GeV. Leptons are counted and the events
classified accordingly ($ee, eee, \mu\mu, e\mu, e\mu\mu$).
Additional electrons are allowed. 
In the $ee$ sample at high invariant mass $M_{ee} > 100$ GeV three events are selected where 0.44 are expected. 

H1 published a measurement of production of $\tau$-leptons\cite{h1taupair}
using 118 pb$^{-1}$ of HERA-I data. 
Elastic $\tau$-pair decays with the final state combinations 
$e$-$\mu$, $e/\mu$-$jet$ ($jet$ = hadronic $\tau$-decay) and
$jet$-$jet$ in the phase space $P_{T\tau} > 2$ GeV and $20^{\circ}<\theta_{\tau}<120^{\circ}$ are analysed.
In the final sample 30 events are observed for $27.1 \pm 4.1$ expected from the SM with a $\tau$-pair contribution of $> 50\%$.
The measured cross section of $13.6 \pm 5.7$ pb is in good agreement with $11.2 \pm
0.3$ pb expected from the SM.

\begin{figure}[thb] 
\centerline{
  \psfig{file=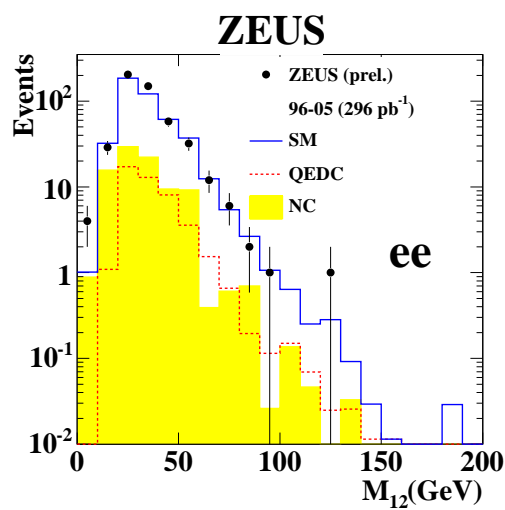,width=65mm,angle=-90}
}
\caption{Invariant mass of the two electrons in the di-electron sample 
observed by the ZEUS experiment. \label{fig:meezeus}}
\end{figure}

A ZEUS preliminary analysis\cite{ref:zeus-mlep} selects multi-electron 
samples ($ee, eee$) and pairs of $\tau$-leptons in a data sample corresponding 
to 296 pb$^{-1}$.
The selection of multi-electron events is similar to the one done by H1. 
Figure \ref{fig:meezeus} shows the invariant mass $M_{ee}$ in the $ee$ sample.
At $M_{ee} > 100$ GeV one event is observed where 1.5 are expected from the SM.
Also $\tau$-pair decays with a $e$-$\mu$ final state were searched in 
elastic events. 
Three events are observed  and $2.0 \pm 0.8$ and expected with negligible background.

All multi-lepton samples in both experiments show good overall agreement with the 
SM prediction, while the events at high mass remain intriguing.

\section{Search for Doubly Charged Higgs Boson} \label{sec:hpp}

\begin{figure*}[thb]
\setlength{\unitlength}{1cm}
\centerline{
  \psfig{file=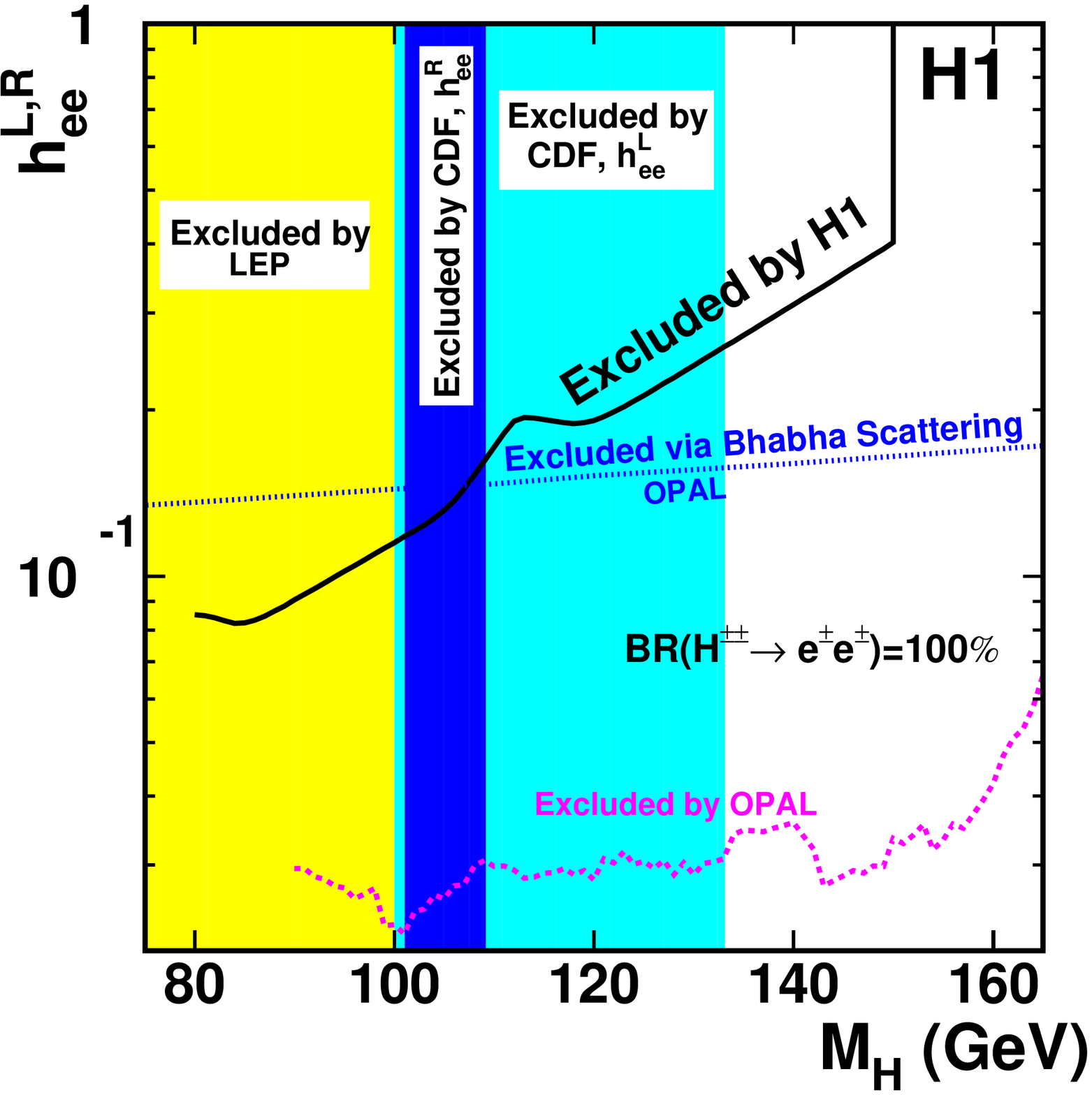,width=50mm} 
  \psfig{file=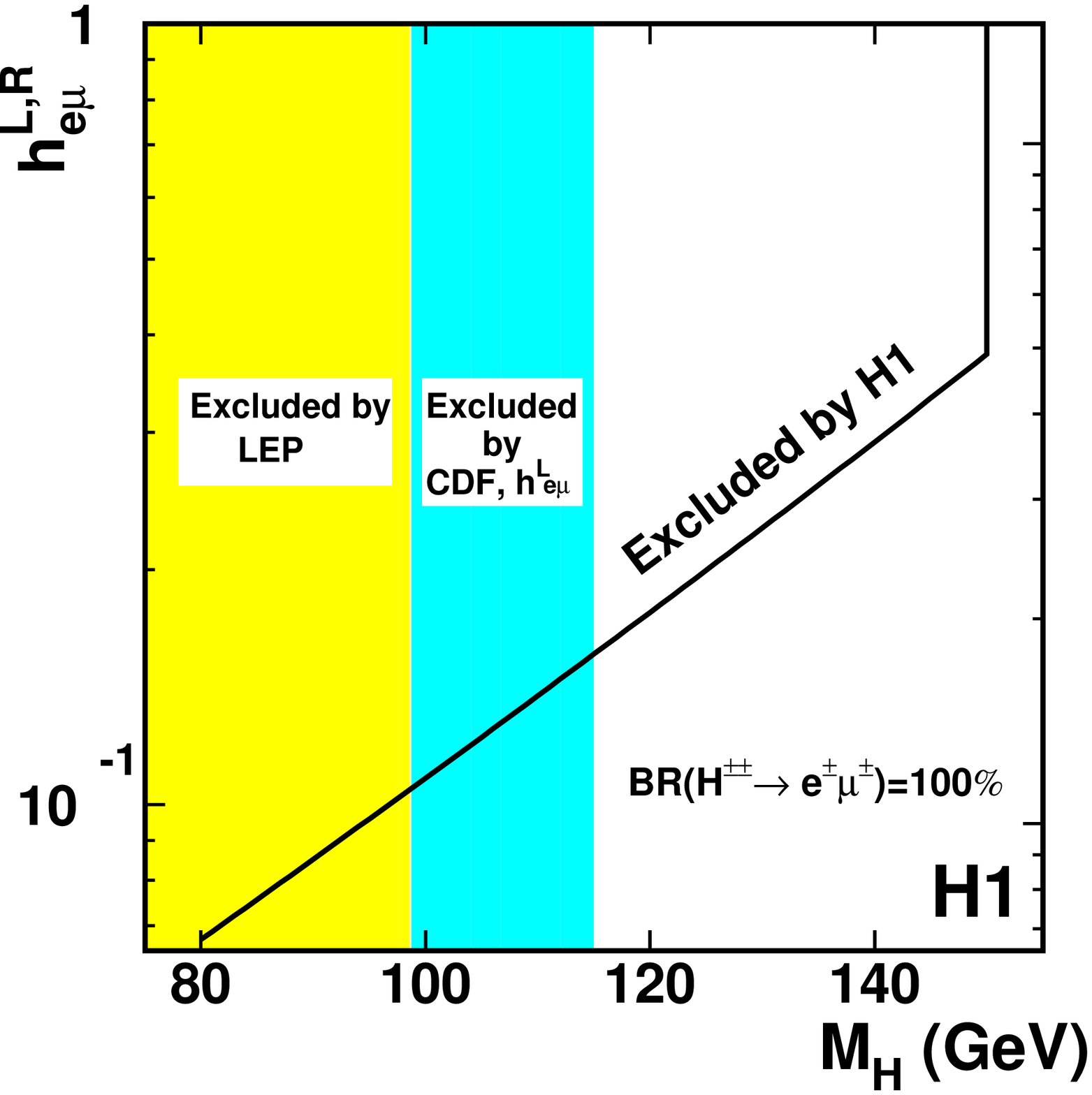,width=50mm} 
  \psfig{file=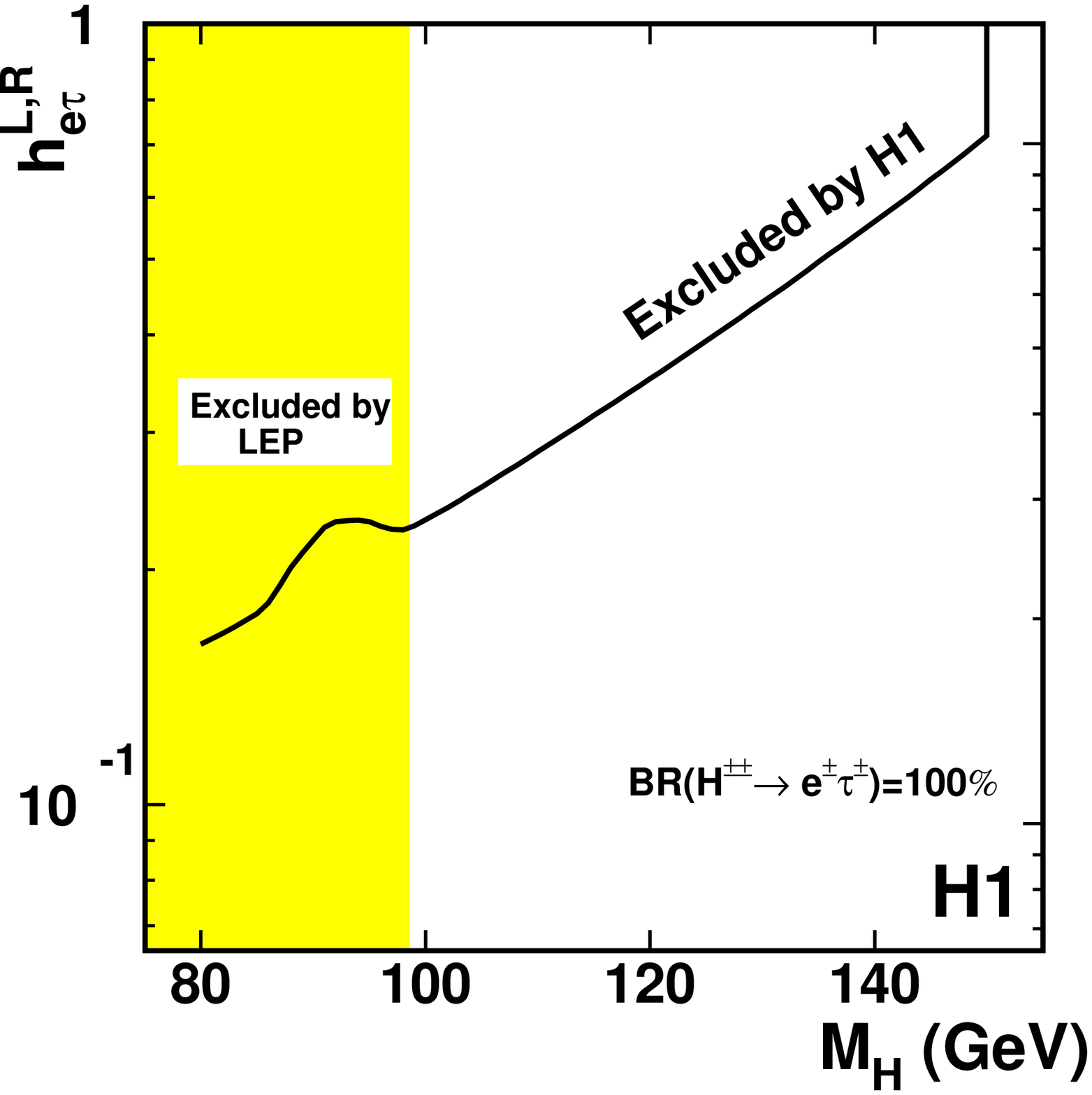,width=50mm}
} 

\caption{Limits on doubly charged Higgs production assuming dominating
couplings $h_{ee}$ (left), $h_{e\mu}$ (middle) or $h_{e\tau}$ (right).}
\label{fig:hpp_limits}
\end{figure*}

The excess of $ee$ and $eee$ events in HERA-I data at H1 has triggered a search for
doubly charged Higgs bosons $H^{\pm\pm}$. 
These appear in various extensions of the SM, eg. left-right symmetric 
models\cite{ref:hpptheo}. 
A $H^{\pm\pm}$ can be radiated off the incoming electron if there are non-vanishing
Yukawa couplings $h_{ee}$, $h_{e\mu}$ or $h_{e\tau}$. 
The final state contains one electron and a lepton pair with the same charge as 
the incident electron. The present analysis\cite{ref:hpp06} investigates
$H^{\pm\pm}$-decays into $ee$, $e\mu$ and $e\tau$ pairs using the unpolarised 
HERA-I data.
The $ee$ and $e\mu$ channels are based on the selection presented in 
Sec. \ref{sec:mlep}.
The $e\tau$ channel is investigated in $\mathcal{L} = 88$ pb$^{-1}$ of HERA-I data
considering all possible decays of the $\tau$ lepton ($e$, $\mu$ and hadronic) in
the phase space $P_T^{e,\tau}>10,5$ GeV and $20^{\circ}<\theta_{e,\tau}<140^{\circ}$.
In the final selection no excess over the SM expectation is observed.
Figure \ref{fig:hpp_limits} shows the upper limits on the 
Yukawa couplings $h_{ee}$, $h_{e\mu}$ and $h_{e\tau}$ obtained by assuming
that one coupling dominates. 
For dominating $h_{e\mu}$, $h_{e\tau}$ couplings H1 extends beyond 
the reach of searches from LEP\cite{ref:hpplep} and TeVatron\cite{ref:hpptev} 
(also shown).
Assuming a coupling of electromagnetic strength $h_{e\mu} (h_{e\tau}) 
= 0.3, M_{H^{\pm\pm}} < 141 (112)$ GeV is excluded.

\section{General Search} \label{sec:generic}

\begin{figure*}[thb]
\setlength{\unitlength}{1cm}
\centerline{
  \psfig{file=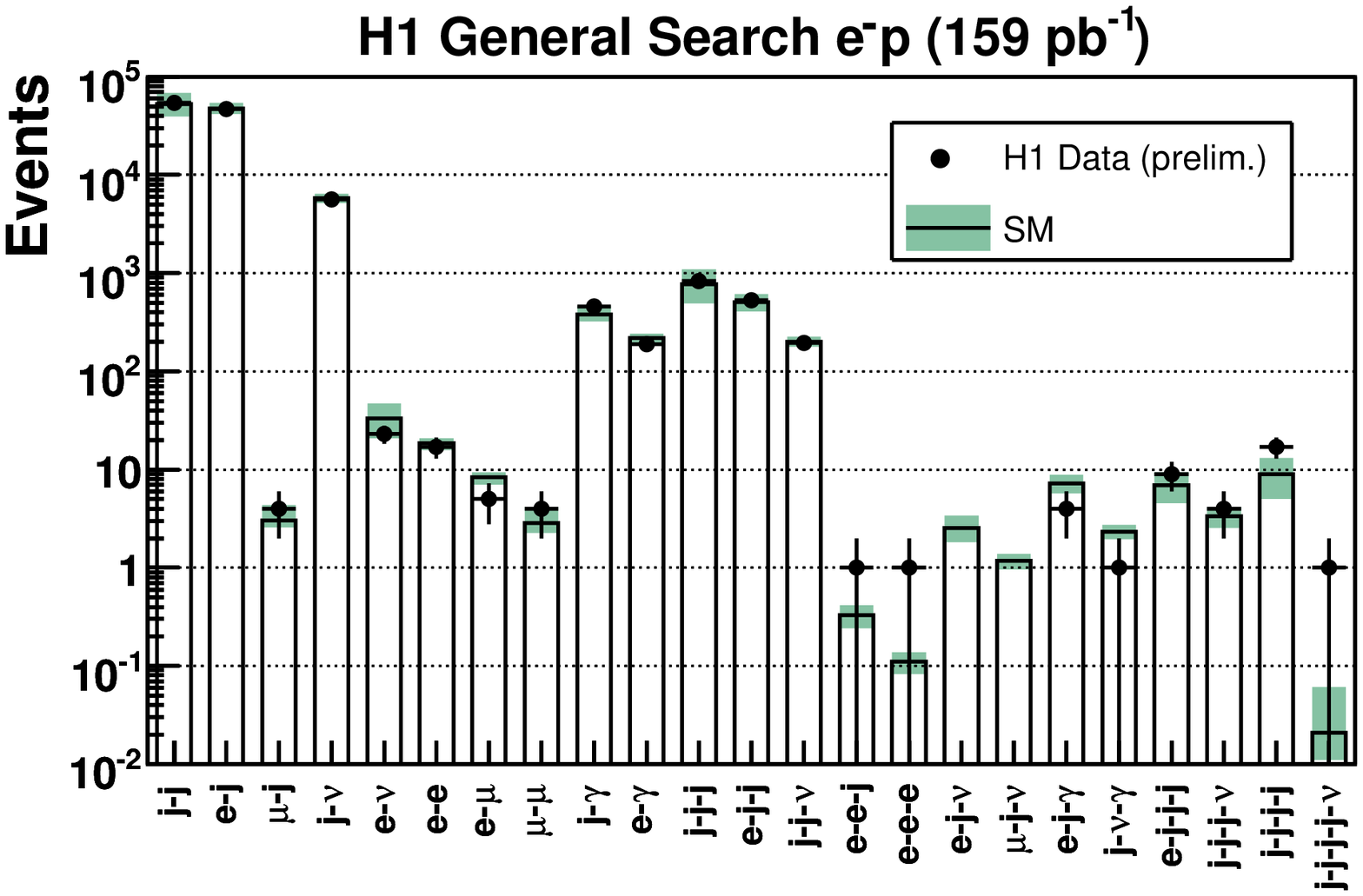,width=120mm}
}
\caption{HERA~II $e^{-}p$ event yields in the general search for new phenomena at 
high $p_T$. Data (points) are compared to the SM expectation
(histogram bars). The uncertainty of the SM prediction is indicated by
the shaded band.}
\label{fig:generic_class1}
\end{figure*}

The H1 collaboration has performed a general search for new high-$P_T$ phenomena by
looking for deviations from the SM\cite{ref:generic2}. 
The HERA-II $e^{-}p$ sample of 159 pb$^{-1}$ is
a complement to the published HERA-I $ep$ sample (118 pb$^{-1}$) dominated by 
87\% of $e^{+}p$ data\cite{ref:generic1}. 
The final states are classified by
the number of identified electrons ($e$), muons ($\mu$), jets ($j$), 
photons ($\gamma$) and missing energy ($\nu$, eg. neutrinos). 
At least two such objects are required. 
They are identified in the phase space $P_T > 20$ GeV and 
$10^{\circ} < \theta < 140^{\circ}$ and have to be isolated.
The results for HERA-II are shown in Fig. \ref{fig:generic_class1}. 
There are 22 channels with at least one observed or
expected event. 
Also differential distributions of the invariant mass $M_\mathrm{all}$ of all objects
in a class and the scalar sum of all transverse momemta $\sum P_T$ were investigated by a statistical algorithm looking for deviations (excess or deficit) from the SM\cite{ref:generic1}. Good agreement with the SM is found in all channels.

\section{Conclusions} \label{sec:conclusion}

The recent HERA-II $e^{-}p$ sample is a great complement to the mostly $e^{+}p$
sample from HERA-I and provides a ten-fold increase in statistics of $e^{-}p$
collisions.
Stricter limits on excited neutrino production could be set
and the reach extended to higher masses. 
This large $e^{-}p$ sample is also used to perform a general search for deviations to the SM in a large variety of high $P_T$ topologies. A good agreement with the predictions from the SM is observed in all channels.

\section*{Acknowledgements} \label{sec:ack}

I thank my colleagues at H1 and ZEUS, in particular C.~Diaconu, E.~Sauvan, 
E.~Perez, M.~Corradi, E.~Tassi, O.~Ota, N.~Okazaki and J.~Ferrando for providing 
me with neccessary materials and usefull comments.

\end{document}